\documentstyle[12pt,epsf,epsfig,wrapfig]{article}
\begin{document}
\begin{center}
{\large \bf
The transverse polarization distribution ${\bf h_1(x,Q^2)}$
\\ }
\vspace{5mm}
\underline{V.~Barone}$^1$, T.~Calarco$^2$ and A.~Drago$^2$
\\
\vspace{5mm}
{\small\it
(1) Universit{\`a} di Torino and INFN, 10125 Torino, Italy \\
(2) Universit{\`a} di Ferrara and INFN, 44100 Ferrara, Italy \\ }
\end{center}

\begin{center}
ABSTRACT

\vspace{5mm}
\begin{minipage}{130 mm}
\small
We report the results of 
a quark model calculation of the transverse polarization
distribution $h_1$. Predictions for the 
Drell-Yan double spin transverse asymmetry are
also presented. 
\end{minipage}
\end{center}

The transverse polarization distribution $h_1(x)$ [1] measures
the polarization asymmetry of quarks in a
transversely polarized hadron, {\it i.e.} the probability
to find a quark polarized in a transverse direction
$+ \hat n$ minus the probability to find it
polarized in the opposite direction $- \hat n$, 
when the
proton's spin points in the direction $+ \hat n$:
\begin{equation}
h_1^q (x) = q_{+ \hat n}(x) - 
q_{- \hat n}(x)\,.
\label{1}
\end{equation}
$h_1(x)$ is a leading twist chirally--odd distribution
function which decouples
from polarized deep inelastic scattering  and is
measurable only in hadron--hadron scattering and in semi-inclusive
reactions. No data on $h_1$ are available at present and very little is
known about it from a theoretical viewpoint.

We calculated $h_1$ in the chiral chromodielectric model (CCDM) [2], 
a relativistic quark model with dynamical confinement, 
successfully used to compute other
distribution functions [3]. This model computation 
provides $h_1$ at a very low scale $Q_0^2$ (our estimate is 
$Q_0^2 \simeq 0.16$ GeV$^2$). 
\begin{figure}[h]
\hspace*{-1truecm}\parbox{8cm}{
\epsfig{figure=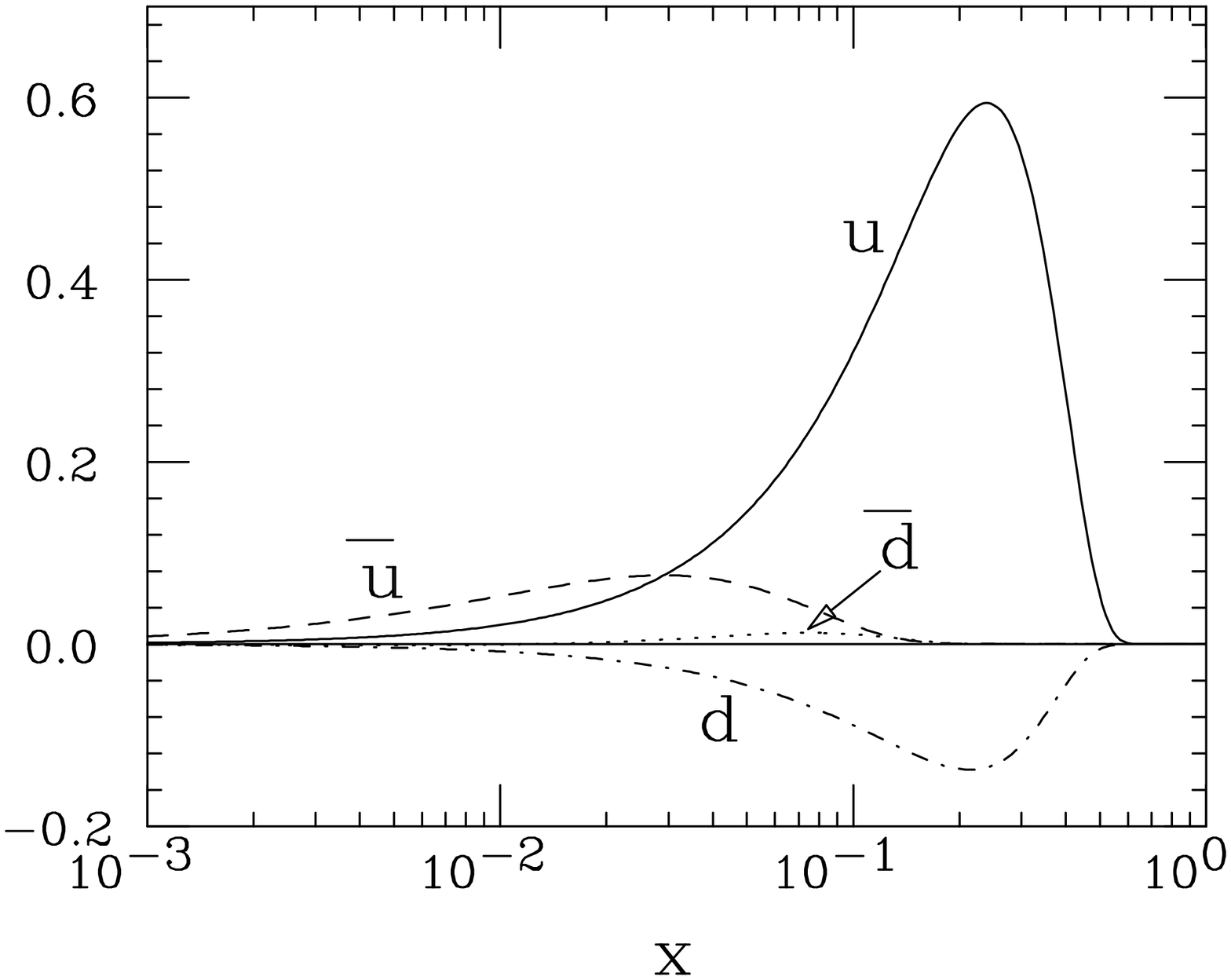,width=8cm,angle=90}
\hspace*{1truecm}\parbox{6.5cm}{\small Figure 1: Flavor
structure of $xh_1(x,Q^2)$ at $Q^2 = 25$ GeV$^2$.\\
\\
}
}
\nolinebreak\parbox{8cm}{
\epsfig{figure=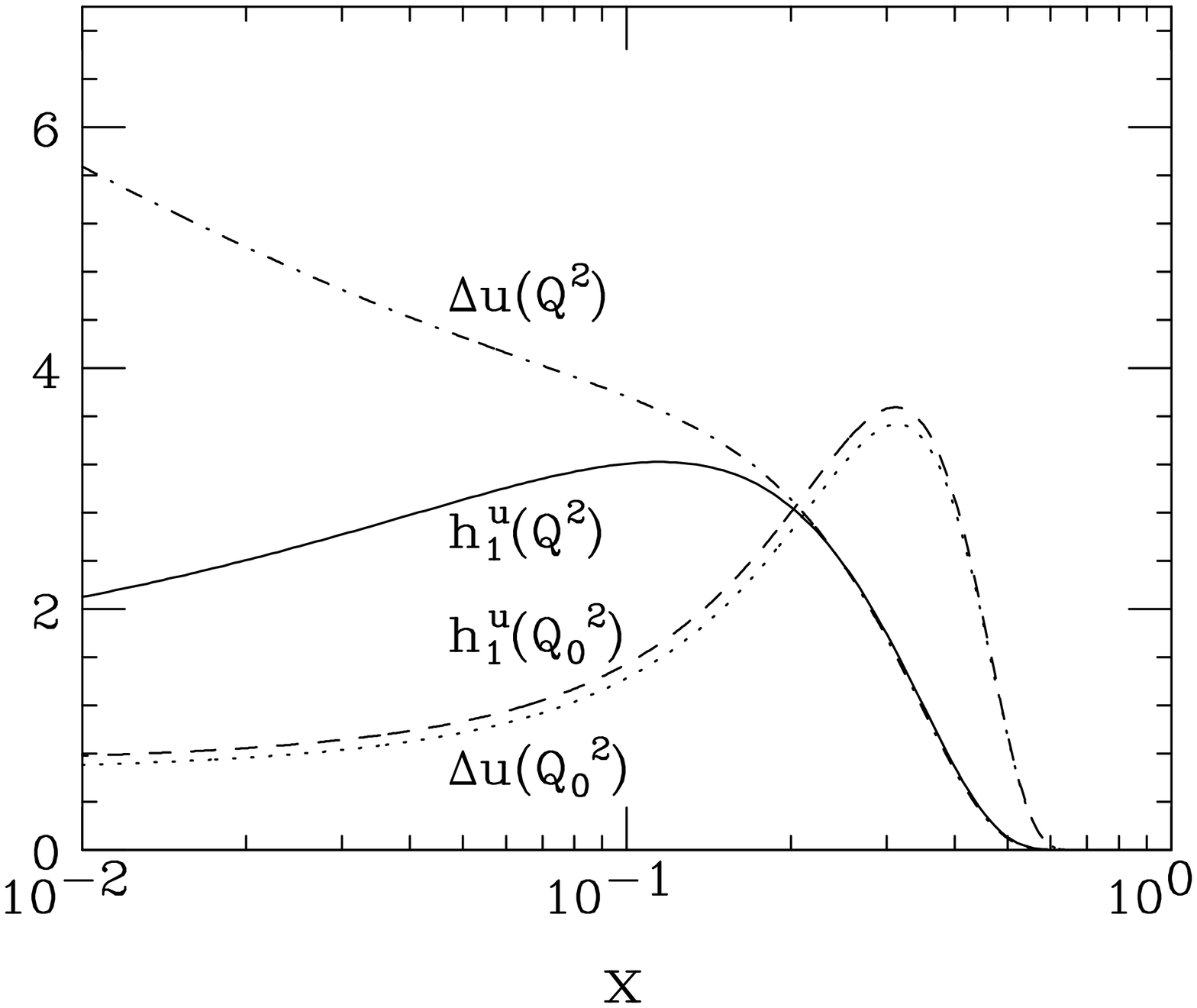,width=8cm,angle=90}
\hspace*{1truecm}\parbox{6.5cm}{\small Figure 2: 
Comparison between transverse and longitudinal polarization 
distributions at $Q_0^2 =0.16$ GeV$^2$ and $Q^2 = 25$ GeV$^2$.}
}
\end{figure}
Starting from $Q_0^2$ the distributions 
are evolved up to larger momentum scales by means of the 
Altarelli--Parisi equation for $h_1$. 
The flavor structure of $h_1$ is presented in fig.~1. In fig.~2 
$h_1^q$ is
compared to its longitudinal counterpart $\Delta q$. 
While at the model scale $h_1^q$ does not
differ much from the helicity distribution $\Delta q$, at larger $Q^2$ 
the
QCD evolution makes these two distributions
considerably different in the small-$x$ region [2,4].  
We also checked that Soffer's relation 
among the three leading--twist 
unpolarized and polarized distribution functions [5] is satisfied in the 
CCDM at the model scale.  
It can be easily shown [4] that if 
Soffer's inequality holds at some scale $Q_0^2$, it will also hold 
at any larger scale, because the Altarelli--Parisi equations 
are structured in such a way to preserve it.

The best way to obtain an  
experimental evidence of $h_1$ is probably  
the Drell--Yan (DY) process with two transversely polarized 
hadron beams. 
In fact, semi-inclusive reactions 
involve unknown twist--3 fragmentation 
and distribution functions,  whereas
other hadron--hadron processes are dominated 
by gluonic diagrams which do not receive contributions from 
the transverse distributions, since there is no analogue 
of $h_1$ for gluons. 
The study of the transverse DY process
is part of the 
physics programme at RHIC. 
The quantity which will be measured is the double spin 
transverse asymmetry $A_{TT}$ which is given at leading order
by ($a$ and $b$ are the two colliding hadrons which we shall suppose
to be both protons)
\begin{equation}
A_{TT} = a_{TT} \, 
\frac{\sum_q h_1^{q}(x_a, M^2) \, h_1^{\bar q}(x_b,M^2) 
+ (a \leftrightarrow b)}{\sum_q q(x_a, M^2) \, \bar q(x_b,M^2) 
+ (a \leftrightarrow b)}\,,
\label{2}
\end{equation}
where $a_{TT}$ is a quantity calculable in perturbative
QCD and varying between -1 and +1, 
$M$ is the mass of the produced dilepton pair, and $x_a,x_b$
are related to the center--of--mass energy $\sqrt{s}$ by 
$x_a x_b = M^2/s$. 
Early calculations of $\vert A_{TT}/a_{TT}\vert$ [6] 
gave values of order 0.1-0.2
for $M=10$ GeV and $\sqrt{s} = 100$ GeV.  
However these predictions relied on the wrong assumption
that $h_1^q= \Delta q$ and $h_1^{\bar q} = \Delta \bar q$ at  all
momentum scales.
\begin{figure}[h]
\hspace*{-1truecm}\parbox{8cm}{
\epsfig{figure=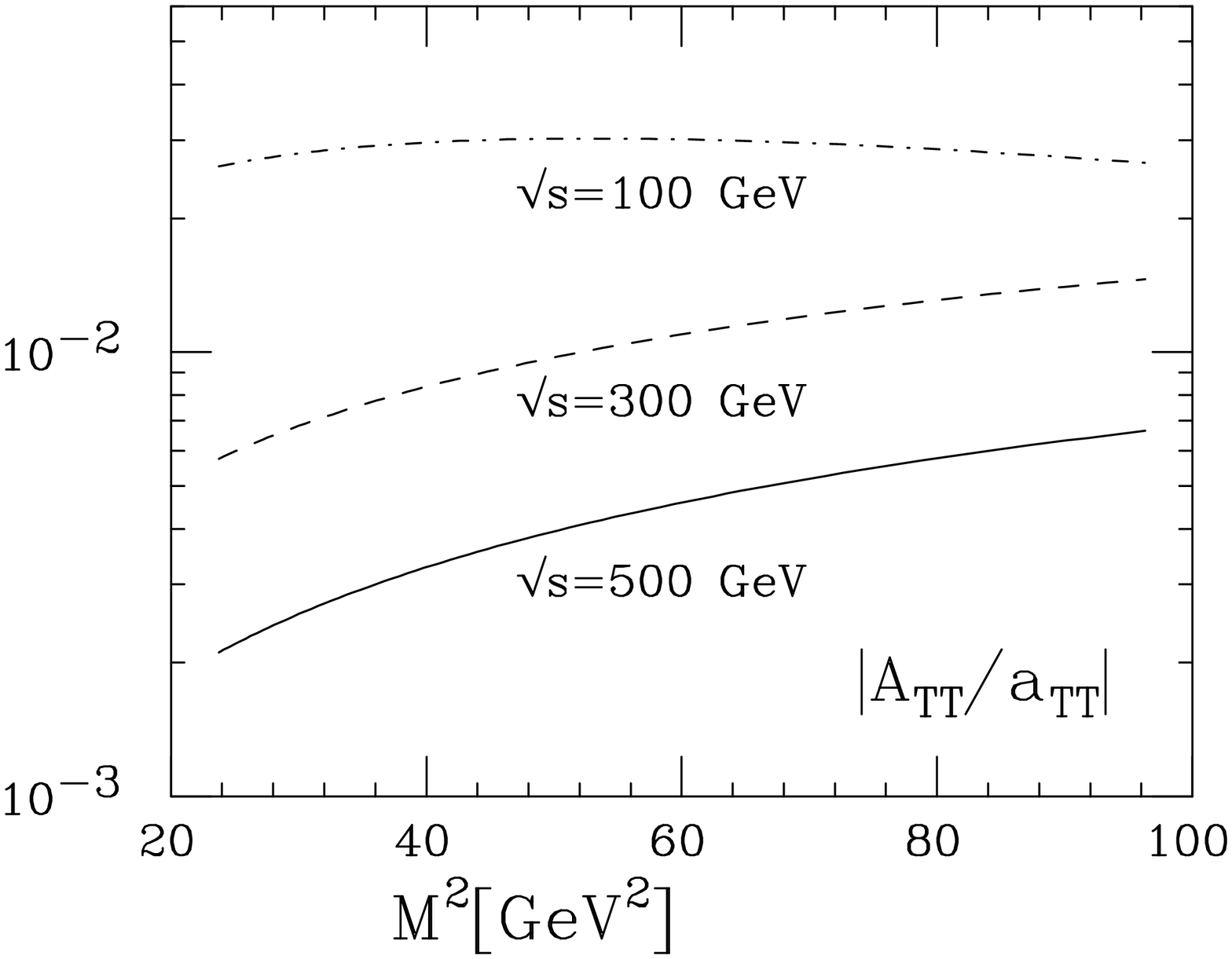,width=8cm,angle=90}
\hspace*{1truecm}\parbox{6.5cm}{\small Figure 3: The 
double transverse asymmetry $\vert A_{TT}/a_{TT} \vert$ at $x_a-x_b=0$, 
computed in 
the chromodielectric model.}
}
\nolinebreak\parbox{8cm}{
\epsfig{figure=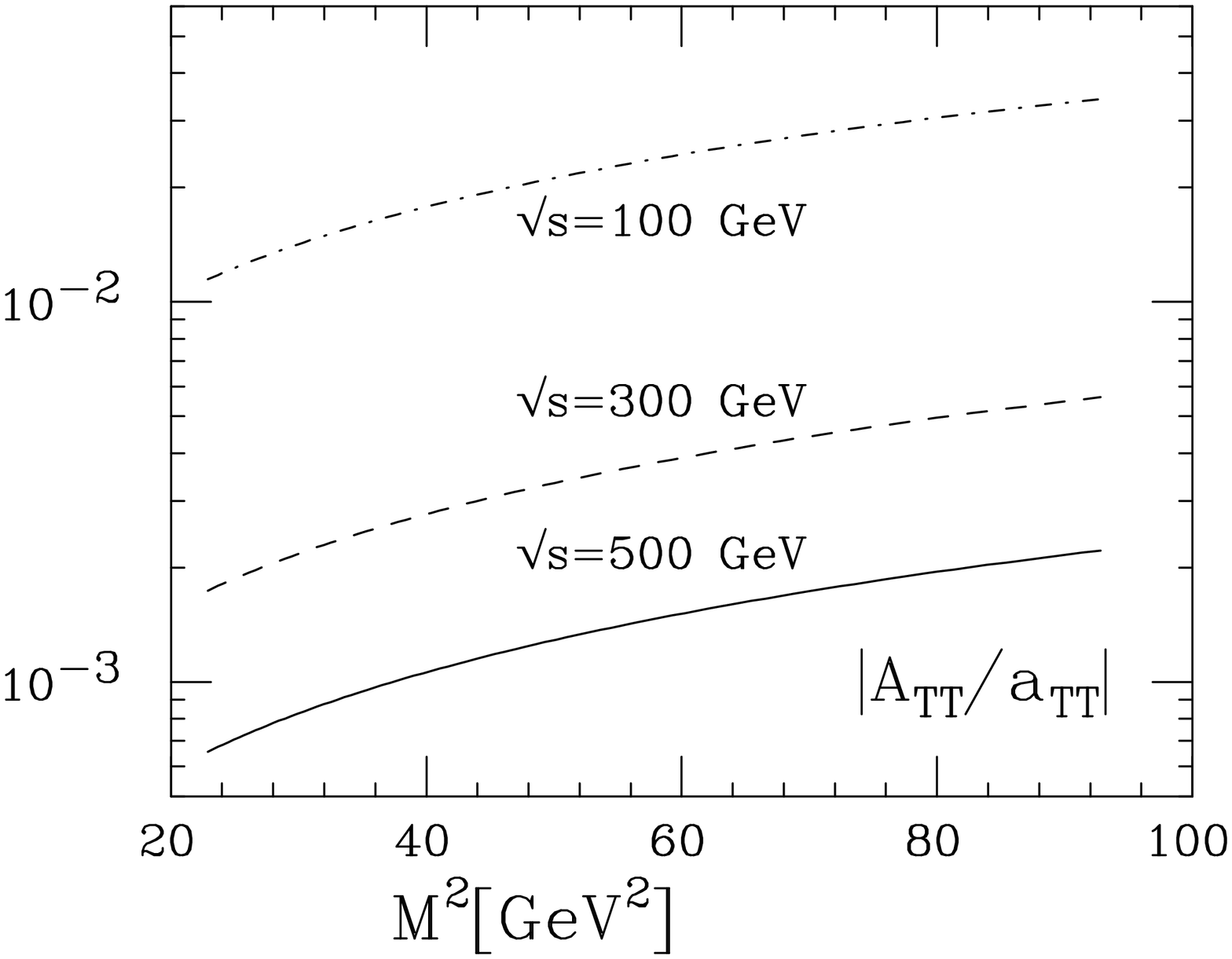,width=8cm,angle=90}
\hspace*{1truecm}\parbox{6.5cm}{\small Figure 4: The 
double transverse asymmetry $\vert A_{TT}/a_{TT} \vert $ at 
$x_a-x_b=0$, computed 
with the GRV parametrization.}
}
\end{figure}

The confinement model results for 
the DY double transverse asymmetry 
are presented in fig.~3  and show that 
$\vert A_{TT}/a_{TT} \vert$ is much smaller 
than it was previously expected.
It also decreases with increasing 
$\sqrt{s}$. 

These results are confirmed by a different calculation 
of $A_{TT}$ based not on a model but on a 
global fit to the longitudinally polarized data [7]. 
We made the realistic hypothesis (motivated by the 
quark model results)
$h_1^q= \Delta q, \; h_1^{\bar q} = \Delta \bar q$   at 
$\mu^2$, 
where $\mu^2$ is a small scale, and used for $h_1$ a 
leading order parametrization 
of the helicity distributions. 
We chose the GRV fit [8], whose input is 
sufficiently low ($\mu^2 = 0.23$ GeV$^2$) to make the above assumption a 
reliable one. 
The results of the 
evaluation of $\vert A_{TT}/a_{TT}\vert $ using the GRV parametrization are 
shown in fig.~4. The GRV 
double transverse asymmetry turns out to 
be of the same order of magnitude as that 
obtained in the chromodielectric model. 
Notice however that 
the sign of the CCDM asymmetry is positive, whereas that 
of the GRV asymmetry is negative, due to the opposite 
signs of the $\Delta \bar u$ densities. 
The fact that two different sets of distributions
yield similar results for the absolute value of $A_{TT}$ 
strengthens our conclusion that the double transverse asymmetry
is at most of order of few percent 
in the typically accessible dynamical regions. 
A larger asymmetry is obtained in 
$p \bar p$ Drell--Yan  production but the experimental investigation of 
this process seems to be beyond  
the present possibilities.

\vspace{0.2cm}
\vfill
{\small\begin{description}

\item{[1]}
J.P.~Ralston and D.E.~Soper, Nucl. Phys. {\bf B152} (1979) 109.
X.~Artru and M.~Mekhfi, Z. Phys. {\bf C45} (1990) 669.
R.L.~Jaffe and X.~Ji, Phys. Rev. Lett. {\bf 67} (1991) 552; 
Nucl. Phys. {\bf B375}
(1992) 527.
J.L.~Cortes, B.~Pire and J.P.~Ralston, Z. Phys. {\bf C55} (1992) 409.

\item{[2]} 
 V.~Barone, T.~Calarco and A.~Drago, 
Torino preprint DFTT 24/96, to be published in Phys. Lett. B.

\item{[3]}
V.~Barone and A.~Drago, Nucl. Phys. {\bf A552} (1993) 479; {\bf A560}
(1993) 1076 (E). 
V.~Barone, A.~Drago and M.~Fiolhais, Phys. Lett. {\bf B338} (1994) 433.

\item{[4]}
V.~Barone, Torino preprint DFTT 68/96. 

\item{[5]}
J.~Soffer, Phys. Rev. Lett. {\bf 74} (1995) 1292.
G.R.~Goldstein, R.L.~Jaffe and X.~Ji, 
Phys. Rev. {\bf D52} (1995) 5006.

\item{[6]}
X.~Ji, Phys. Lett. {\bf B284} (1992) 137. 
C.~Bourrely and J.~Soffer, Nucl. Phys. {\bf B423} (1994) 329.

\item{[7]}
V.~Barone, T.~Calarco and A.~Drago, paper in preparation. 

\item{[8]}
M.~Gl{\"u}ck, E.~Reya and W.~Vogelsang, Phys. Lett. {\bf B359} (1995) 201.

\end{description}}

\end{document}